\documentclass{aa}
\usepackage{graphics,epsfig}
\newcommand{\simgeq}{\ensuremath{\widetilde{>}}}
\newcommand{\kms}{km~s$^{-1}$}

\begin{document}

\title{Kinematics of the dwarf irregular galaxy GR8}
\titlerunning{Kinematics of GR8}
\author{Ayesha Begum\inst{1}\thanks{ayesha@ncra.tifr.res.in} and 
        Jayaram N. Chengalur\inst{1}\thanks{chengalur@ncra.tifr.res.in}
}
\authorrunning{Begum \& Chengalur}
\institute{National Centre for Radio Astrophysics,
	Post Bag 3, Ganeshkhind, Pune 411 007}
\date{Received mmddyy/ accepted mmddyy}
\offprints{Ayesha Begum}
\abstract{
	 We present deep, high velocity resolution ($\sim~1.6$~\kms)
Giant Meterwave Radio Telescope HI 21cm synthesis images for the faint
($M_B \sim -12.1$) dwarf irregular galaxy GR8. We find that the velocity 
field of the galaxy shows a clear systematic large scale pattern, with 
a maximum amplitude $\sim 10$~\kms. Neither pure rotation, nor pure 
radial motion alone can fit the observed velocity field; however a 
combination of radial and circular motions can provide a reasonable
fit. The most natural interpretation is that the neutral ISM, in addition
to rotating about the center, is also expanding outwards, as a result of
energy input from the ongoing star formation in the galaxy. Support 
for this interpretation comes from the fact that the pressure in the 
HII regions in the galaxy is known to be substantially ($\sim 55$ times)
more than the average pressure in the gas disk. It is, however, also 
possible that the velocity field is the result of the gas swirling 
inwards, in which case GR8 could be in the process of formation via 
the merger of subgalactic clumps.  
}
\maketitle
\section{Introduction}
\label{intro}

	Although bright irregular galaxies have rotating gas disks, 
it is unclear whether the faintest dwarf irregular galaxies are 
rotationally supported or not. C\^{o}t\'{e} et al. (2000), based on a 
study of the kinematics of eight dwarf irregular galaxies (with magnitudes
varying from M$_{\rm B} = -16.7$ to M$_{\rm B} = -11.3$) suggest that
normal rotation is seen  only in dwarfs brighter than M$_B \sim -14$.
This is consistent with the earlier findings of \cite{lo93}, who, 
based on an interferometric study of faint dwarf galaxies (with M$_{B\rm} 
\sim -9$ to M$_{B\rm} \sim -15$) found that only two of their sample
of nine galaxies showed ordered velocity fields. However, this conclusion
has been questioned by \cite{skillman96} who pointed out that the 
interferometric observations of \cite{lo93} lacked sensitivity to 
low extended HI distribution and could thus have been insensitive to
the large scale velocity field. Further, \cite{begum03} showed that 
the dwarf irregular galaxy Camelopardalis~B, despite being extremely
faint ($M_B \sim -10.9$) nonetheless has a regular velocity field, 
consistent with that expected from a rotating disk. So, some faint
galaxies at least, have rotating HI disks. What about the rest? If gas in 
faint dwarf galaxies is not supported by rotation, what provides the energy 
that keeps it from collapse? For very faint dwarf irregular galaxies, the
binding energy of the gas is not much larger than the energy output of a few
supernovae. Star formation in such galaxies could hence have a profound 
effect on the kinematics of the ISM. Indeed, the faintest dwarf galaxies are 
expected to lose a substantial part of their gas due to the energy 
deposited in the ISM by supernovae from the first burst of star formation 
(e.g. Dekel \& Silk 1986). Some observational support for such models
is provided by the large expanding  HI supershells seen in the ISM of
some dwarf irregular galaxies with active star formation (e.g. 
Ott et al. 2001). In this paper we discuss the issue of the kinematics
of faint dwarf galaxies, and its possible connections with energy
input from stellar processes, in the specific context of the 
faint ($M_B \sim -12.1$) dwarf irregular galaxy GR8.

	GR8 was first discovered by \cite{reaves56} in the course of a survey
for dwarf galaxies in the direction of the Virgo Cluster. It has also 
been cataloged as DDO~155 by \cite{vdberg59}. The original distance 
estimates for GR8 were in the range 1.0 - 1.4~Mpc (\cite{hodge67,devac83,
hoessel83}), which would make GR8 a probable member of the local group. 
However, recent estimates give somewhat larger distances. 
\cite{tolstoy95} estimated a distance of 2.2 Mpc (based on observations of
the only detected Cepheid variable). This estimate is in excellent 
agreement with that of \cite{dohm98} which is based on the brightness 
of the tip of the red giant branch. From the location for the local 
group barycenter given by \cite{courteau99} one can  compute that 
this distance places GR8 well outside the local group zero velocity 
surface. Consistent with this, \cite{vdberg00} does not classify 
GR8 as a member of the local group.

	GR8 has a patchy appearance in optical images, with the emission
being dominated by bright blue knots. H-$\alpha$ imaging (Hodge 1967)
shows that these knots are sites of active star formation. However, in 
addition to the bright blue knots, GR8 also possesses faint extended 
emission (\cite{hodge67, devac83}), indicative of earlier episodes
of star formation. Indeed, CM diagrams (based on HST imaging, 
\cite{dohm98}), show that although the bright star forming knots in 
GR8 have stars which are younger than $\sim 10$~Myr, the galaxy also
contains stars which are older than a few Gyr. Despite 
this long history of star formation, the metallicity of the star
forming knots in GR8 is extremely low, $\sim 3\%$ solar (Skillman 
et al. 1988b). This makes it one of the lowest metallicity galaxies known 
(Kunth \& \"{O}stlin 2000). In keeping with this low metallicity, despite 
the fact that the star forming regions are expected to be associated with
molecular gas, CO has not been detected in the galaxy (Verter \& Hodge 1995).

	There have been two independent studies of the kinematics of  HI 
in GR8,  both using the VLA. However these two studies resulted in very 
different interpretations of the galaxy's kinematics. \cite{carignan90} 
assumed that the observed velocity field was produced by rotation
and used it to derive a rotation curve. On the other hand \cite{lo93} 
interpreted the velocity field as being due to radial motions (i.e. either 
expansion or contraction). Both of these studies were based on modest 
($\sim 6$~\kms) velocity resolution observations. Further both observations
used the VLA C array, and hence were not sensitive to emission from the
extended low surface brightness portions of the HI disk. There has also 
been a recent high velocity resolution VLA (Cs array) based study of 
GR8 (Young et al. 2003). This study was focused on the local connections 
between the ISM and star formation and not the large scale kinematics
of the gas. Although \cite{young03} noted that velocity field in GR8 does 
show a large scale gradient, they chose to characterize the velocity
field as giving the overall impression of resulting from random
motions.

	We present here deep, high velocity resolution ($\sim~1.6$~\kms) 
Giant Meterwave Radio Telescope (GMRT) observations of the HI emission 
from GR8 and use them to study the kinematics of this galaxy.
The rest of the paper is divided as follows. The GMRT observations 
are detailed in Sect.~\ref{sec:obs}, while the results are presented
in discussed in Sect.~\ref{sec:res}. Throughout this paper we take 
the distance to GR8 to be 2.2 Mpc, and hence its absolute magnitude to 
be $M_B \sim -12.1$.

\section{Observations}
\label{sec:obs}     
    
    The GMRT observations of GR8 were conducted from 16$-$18 November
2002. The setup for the  observations is given  in Table~\ref{tab:obs}.
Absolute flux calibration was done using scans on the standard calibrators
3C48 and  3C286,  one of which was observed at the start and end of 
each observing run. Phase calibration was done using 1252+119 which 
was observed once every 30 minutes. Bandpass calibration was done in 
the standard way using 3C286. 

     The data were reduced using standard tasks in classic AIPS.  For each run,
bad visibility points were edited out, after which the data were calibrated.
Calibrated data for all runs were combined using DBCON. The GMRT does not
do online doppler tracking -- any required doppler shifts have to be applied
during the offline analysis. However since the differential doppler shift
over our observing interval is much less than the channel width, there was
no need to apply an offline correction.

     The GMRT has a hybrid configuration (Swarup et al. 1991) with 14 of its
30 antennas located in a central compact array with size $\approx$ 1 km 
($\approx$ 5 k$\lambda$ at 21cm) and  the remaining antennas distributed 
in a roughly ``Y'' shaped configuration, giving a maximum baseline length 
of $\approx$ 25 km ($\approx$ 120 k$\lambda$ at 21 cm). The baselines 
obtained from antennas in the central square are similar in length to 
those of the ``D'' array of the VLA while the baselines between the arm 
antennas are comparable in length to the ``B'' array of the VLA. A single 
observation with the GMRT hence yields information on both large and small
angular scales. Data cubes were therefore made at various (u,v) ranges, 
including 0$-$5 k$\lambda$, 0$-$10 k$\lambda$, 0$-15$ k$\lambda$ and 0$-$80  
k$\lambda$ using uniform weighting. At each (u,v) range,  a circularly 
symmetric gaussian taper with a FWHM equal to 80\% of the (u,v) range was 
applied, in order to reduce the sidelobes of the synthesized beam. The 
angular resolutions obtained  for the  various (u,v) ranges  listed above 
were 40$''\times38''$, 25$''\times25''$, 16$''\times14''$ and 4$''\times3''$
respectively. All the three low resolution data cubes (i.e. up to 16$''\times14''$ resolution)
were deconvolved using the the AIPS task IMAGR. For the highest resolution
data cube, the signal to noise ratio was too low for CLEAN to work 
reliably and hence the 4$''\times3''$ resolution data cube could 
not be deconvolved. Despite this, the low SNR of this image implies that 
the inability to deconvolve it does not greatly degrade its dynamic range
or fidelity. The morphology of the emission should hence be accurately 
traced, apart from an uncertainty in the scaling factor (this essentially
arises because the main effect of deconvolving weak emission at about the
noise level corresponds to multiplying by a scale factor; 
\cite{jorsater95,rupen99}).

    The HI emission  from  GR8 spanned  28 channels of  the spectral cube. 
A continuum image was made using the average of remaining line free
channels. No continuum was detected from the disk of GR8 to a $3\sigma$
flux limit  of 1.8~mJy/Bm (for a beam size of $46^{''}\times37^{''}$).
We also checked for the presence of any compact continuum sources in 
the disk of GR8 by making a high resolution ($6.4^{''}\times5.8^{''}$)
map --  no sources associated with the disk of GR8 were detected down
to a $3\sigma$ limit of 0.6~mJy/Bm.

        We examined the line profiles at various locations in the galaxy
and found that they were (to zeroth order) symmetric and  single peaked.
In the very high column density regions, a double gaussian and/or
a gauss-hermite fit does provide a somewhat better description of the data,
but even in these regions, the mean velocity produced by the moment 
method agrees within the errors with the peak velocity of the profile.
Since we are interested here mainly in the systematic velocities, moment
maps provide an adequate description of the data.  Moment maps (i.e. 
maps of the total integrated flux (moment~0),  the flux weighted velocity
(moment~1) and the flux weighted velocity dispersion (moment~2)) were 
made from the data cubes using  the AIPS task MOMNT. To obtain the moment
maps, lines of sight with a low signal to noise ratio were excluded by 
applying a cutoff at the $3\sigma$ level, ($\sigma$ being the rms noise level
in a line free channel), after smoothing in velocity (using boxcar
smoothing three channels wide) and position (using a gaussian with
FWHM $\sim 2$ times that of the synthesized beam). Maps of the velocity
field and the velocity dispersion were also made in GIPSY using single 
gaussian fits to the individual profiles. The velocities produced by 
MOMNT in AIPS are in reasonable agreement with those obtained using a single 
gaussian fit. However the AIPS moment~2 map systematically underestimates
the velocity dispersion (as obtained from gaussian fitting) particularly
near the edges where the signal to noise ratio is low. This can be 
understood as the effect of the thresholding algorithm used by the MOMNT
task to identify the regions with signal. From the gaussian fitting we find
that the velocity dispersion $\sigma$ is $\approx$ 9.0 km sec$^{-1}$,
and shows only slight variation across the galaxy. This value of 
$\sigma$ and the lack of substantial variation of $\sigma$  across the 
galaxy is typical of dwarf galaxies (e.g. \cite{lake90,skillman88a}).

\begin{table}
\caption{Parameters of the GMRT observations}
\label{tab:obs}
\vskip 0.1in
\begin{tabular}{ll}
\hline
Parameters& Value \\
\hline
\hline
RA(1950) &12$^h$56$^m$10.5$^s$\\
Declination(1950) &+${14}^{\circ} 29' 17''$\\
Central velocity (heliocentric) &215.0 km sec$^{-1}$\\
Date of observations &16$-$18 Nov 2002\\
Time on source &16 hrs\\
Total bandwidth &1.0 MHz (211~\kms)\\
Number of channels &128\\
Channel separation &1.65 km sec$^{-1}$\\
FWHM of synthesized beam  &41$^{''}\times39^{''}$, 25$^{''}\times25^{''}$,\\
                          & 4$''\times3''$\\
RMS noise per channel &2.0~mJy, 1.6 mJy,\\ 
                      &1.0 mJy\\
\hline
\end{tabular}
\end{table}

\section{Results and Discussion}
\label{sec:res}
\subsection{HI distribution}
\label{ssec:HI_dis}

\begin{figure}[h!]
\epsfig{file=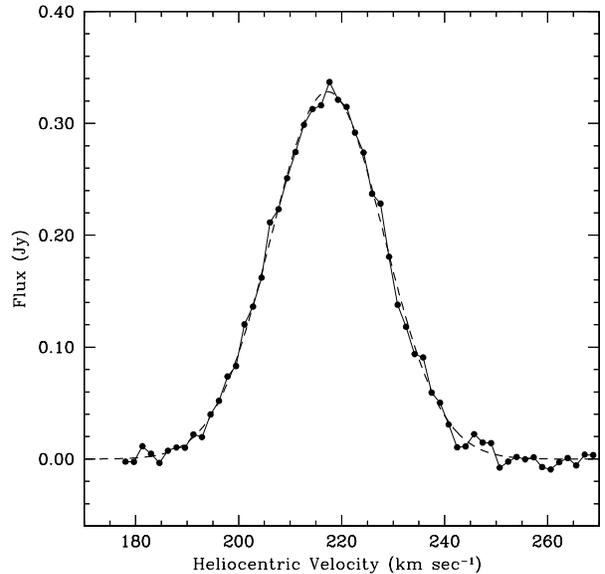,width=3.4in}
\caption{ The integrated spectrum for GR8 obtained from the 40$''\times38''$ 
	  data cube. The channel separation is $1.65$~\kms. Integration of 
	  the profile gives a flux integral of  $9.0$~Jy \kms and an HI 
          mass of $\sim 10.3\times{10}^{6} M_\odot$. The dashed line 
	  shows a gaussian fit to the profile.
         }
\label{fig:HI_spec}
\end{figure}

   The global HI emission profile of GR8, obtained from 40$''\times38''$ 
data cube, is shown in Fig.~\ref{fig:HI_spec}. A Gaussian fit to the 
profile gives a central velocity (heliocentric) of  $217 \pm 2$~\kms.
The integrated flux is $9.0\pm0.9$~Jy~\kms. These are in good agreement
with the values of $214 \pm 1 $~\kms and  $8.78$~Jy~\kms obtained
from single dish observations (Huchtmeier et al. 2000). The good agreement
between the GMRT flux and the single dish flux shows that no flux was
missed because of the missing short spacings in the interferometric
observation. The velocity width at the 50 \% level ($\Delta V_{50}$) is
$26 \pm 1$~\kms, which again is in good agreement with the 
$\Delta V_{50}$ value of $27$~\kms determined from the single dish 
observations. The HI mass obtained from the integrated profile (taking
the  distance to the galaxy to be 2.2~Mpc) is $10.3\pm1.0 \times{10}^{6}
M_\odot$, and the $M_{HI}/L_B$ ratio is found to be $\sim 1.0$ in
solar units.
       
\begin{figure}[h!]
\epsfig{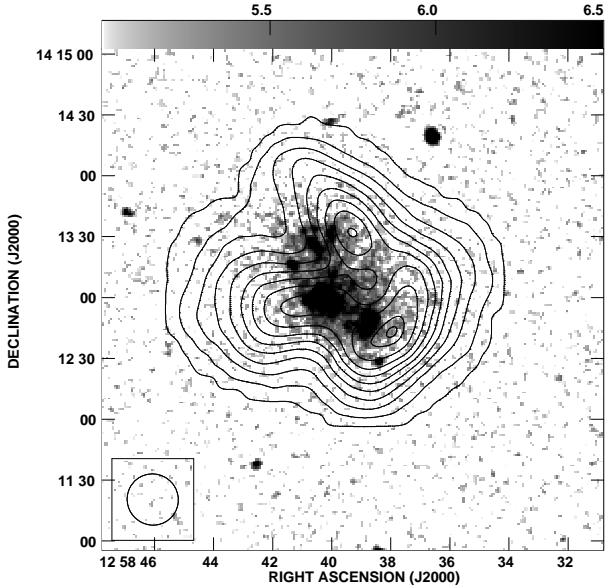}
\caption{The digitized Palomar Sky Survey image of GR8 (greyscales) with 
         the GMRT $25^{''} \times 25^{''}$ resolution integrated HI
         emission map (contours) overlayed. The contour levels are
         0.005, 0.076, 0.146, 0.217, 0.288, 0.359, 0.429, 0.500, 0.571, 0.642
         and 0.665 Jy/Bm~\kms} 
\label{fig:ov}
\end{figure}

\begin{figure}[]
\epsfig{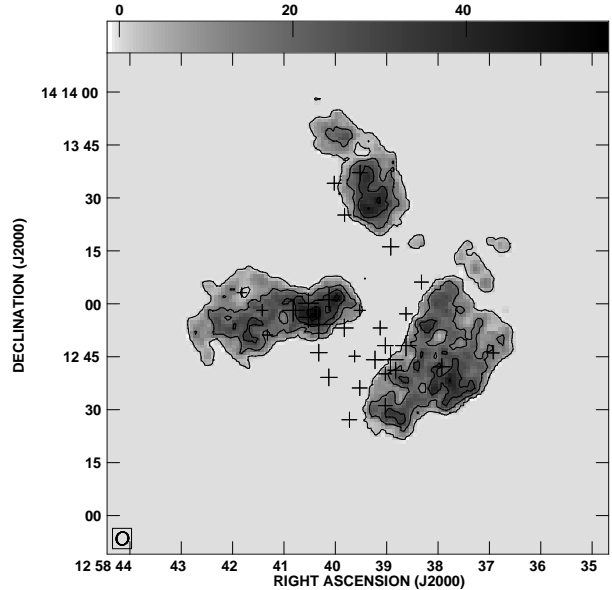}
\caption{ Integrated HI emission at 4$''\times3''$ resolution, (grey scales
          and contours). The contour levels are 0.002, 0.016, 0.030, 0.044 and
          0.058 Jy/Bm~\kms. The locations of the HII regions identified by
          \cite{hodge89} are indicated by crosses.  }
\label{fig:hii}
\end{figure}

	Fig.~\ref{fig:ov} shows the integrated HI emission from GR8  at 
$25^{''}\times25^{''}$~resolution, overlayed on the digitized sky survey 
(DSS)  image. The  HI distribution is clumpy and shows three major  
clumps. This is highlighted in Fig.~\ref{fig:hii} which shows the integrated
HI emission at high resolution ($4.0^{''}\times3.0^{''}$ ). The faint 
extended HI gas seen in the low resolution image is resolved out in this 
image. One may suspect that the diffuse HI emission (particularly that
seen in between the three clumps in the low resolution map) is not real 
but is the result of beam smearing. To check for this possibility,
the individual channel maps in the $25^{''}\times25^{''}$ data cube were 
inspected. In the channel maps, the peak of the diffuse emission in the 
central region of the galaxy occurs at a different heliocentric velocity 
than peak velocities of nearby HI clumps, contrary to what one would expect 
from beam smearing. As a further confirmation of this, the clean components 
from the $25^{''}\times25^{''}$ resolution data cube were  convolved with a 
smaller restoring beam of $10^{''}\times10^{''}$,  to generate a new data 
cube. The diffuse emission is visible in the channel maps in this cube, 
contrary to what would have been expected in case the diffuse emission 
was entirely due to beam smearing (in which case the clean components 
would have been restricted to the three clumps).

	As can be seen in Fig.~\ref{fig:ov}, each  HI clump 
is associated with a clump of optical emission. However, for each clump,
the peak optical emission is generally offset from the peak of the HI 
emission. The H$\alpha$ image of \cite{hodge89} shows that 
the optical clumps also emit copious amounts of H$\alpha$ and are 
hence regions of on going star formation. In addition to the bright
clumps, diffuse optical emission is also seen in Fig.~\ref{fig:ov}.
The optical emission has a much higher ellipticity than the HI emission
and the position angles of the optical and HI major axis can also be seen
to be different. Quantitatively, ellipse fitting to the outermost
contours of the 40$''\times38''$ and 25$^{''}\times25^{''}$ resolution
HI moment maps (which are less distorted by the presence of the 
HI clumps in the inner regions) gives a position angle of 
77$\pm$5 degrees and an inclination (assuming the intrinsic shape
of the HI disk to be circular) of 28$\pm$3 degrees. The values 
obtained from the two different resolution maps agree to within the
error bars. On the other hand, these values are considerably different from 
those obtained from ellipse fitting to the optical isophotes, which 
yields a position angle of 38.4 degrees and an inclination of 57.7 
degrees respectively (De Vaucouleurs \& Moss 1983). We return to this
issue in Sect.~\ref{ssec:discuss}.

\subsection{HI Kinematics}
\label{ssec:HI_Kin}

\begin{figure}[]
\rotatebox{-90}{\epsfig{file=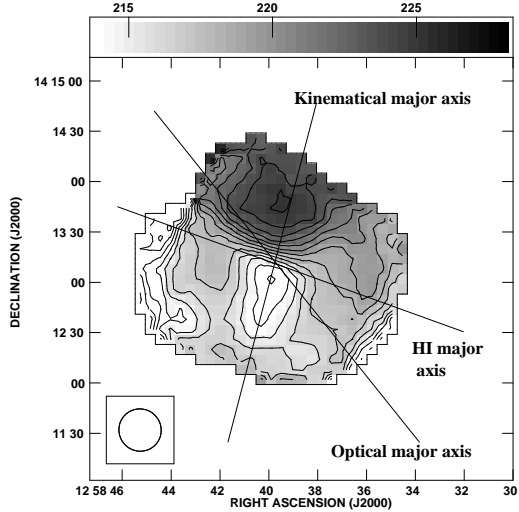,width=3.0in}}
\caption{The HI velocity field of GR8 at $25^{''}\times 25^{''}$ arcsec
          resolution. The contours are in steps of 1~km~sec$^{-1}$ and
          range from 210.0~km sec$^{-1}$ to 225.0~km~sec$^{-1}$. 
        }
\label{fig:mom1}
\end{figure}

	 The velocity field derived from the $25^{''}\times 25^{''}$ 
resolution data cube is shown in Fig.~\ref{fig:mom1}. This velocity field
is in reasonable agreement (albeit of better quality) with that obtained
by \cite{carignan90}. The velocity field shows closed contours and is, 
to zeroth order, consistent with a velocity field that would be 
produced by a rotating disk with an approximately north south kinematical
major axis. This would make the kinematical major axis roughly 
perpendicular to the major axis obtained from ellipse fitting to 
the HI disk. The kinematical major axis is also substantially misaligned 
with the major axis obtained by ellipse fitting to the optical isophotes. 
In addition to this misalignment, the kinematical center of the velocity 
field is offset (to the north, as can be seen by comparing 
Figs.~\ref{fig:mom1} and \ref{fig:ov}) from the center (as determined by
ellipse fitting) of the HI disk.

	Apart from the misalignments mentioned above, the velocity field of
GR8  also shows clear departures from what would be expected from an 
axisymmetric rotating disk. The most important departure is that the 
isovelocity contours in the outer regions of the galaxy show large scale 
kinks. In addition, the velocity field shows several asymmetries. The most 
prominent asymmetry is  between the northern and southern half of the galaxy. 
The closed isovelocity contours in the southern  half are more elongated than
those in the northern half. Further, the kinks noted above are much more 
prominent in the western part of the disk than in the eastern half. Since 
our velocity field is better sampled compared to the velocity fields derived 
by  \cite{lo93} and \cite{carignan90} these kinematical peculiarities are 
more clearly seen. In particular, the offset between the morphological and 
kinematical center, which is apparent  in our velocity field is not seen 
in velocity fields derived earlier.  Further, because of the 
lower sensitivity, the kinks in the isovelocity contours  seen towards
the edges of the galaxy are not seen that clearly in  the earlier
velocity fields. 

	Following \cite{carignan90} we could try to fit GR8's velocity
field to that expected from a rotating disk. In such a fit one can 
anticipate (based on the closed isovelocity contours) that the rotation 
curve would be falling and (based on the kinks in the isovelocity curves
on the eastern and western edges of the disk) that either
the rotation curve would need to rise again towards the edge of the
disk, or the edges of the disk would need to be warped. We discuss 
rotation and other models for producing the observed velocity field 
in more detail in the next section. 

\subsection{Discussion}
\label{ssec:discuss}

	As described in the last two sections, the morphology and kinematics
of GR8 are somewhat peculiar. If the HI gas and the stars in GR8 are both
in disks, then the stellar disk would have to be both more inclined and have
a different position angle than the gas disk. It is more likely that the
star formation in GR8 has occurred preferentially in a non axisymmetric
symmetric region in the center of the galaxy. In the extreme case, the
stars would have a more bar like distribution than the gas. A central
stellar bar could affect the gas dynamics, however
since the stellar mass is probably not dynamically dominant (from the 
observed B-V color of 0.38 for GR8 and the low metallicity models 
of Bell $\&$ de Jong (2001), the the stellar mass is 
$\sim$ 5$\times 10^6$~M$_\odot$, i.e. a factor of 2 less than the HI
mass) this effect may not be important.

   Apart from having a peculiar morphology, the kinematics of GR8 is 
also unusual. The kinematical  and HI major axis of this galaxy are 
perpendicular to each other, the kinematical center is offset from
the morphological center and the observed velocity field is systematically
asymmetric. GR8 is not the only dwarf galaxy which shows misalignment
between kinematical  and morphological axes, such  misalignments have
also  been seen in, for e.g. Sextans~A (Skillman et al. 1988a), NGC~625 
(C\^{o}t\'{e} et al. 2000) and DDO~26 (Hunter \&  Wilcots 2002). However, the
misalignment and off-centered kinematics does imply that GR8 cannot be 
modeled as a pure axisymmetric rotating disk (for which all axis
and centers would be aligned).  Although, \cite{carignan90} had noted
some of these problems, they had nonetheless, modeled the kinematics of GR8 
as an axisymmetric rotating  disk. Their derived rotation curve had a 
maximum amplitude of $\sim 8$~\kms, and fell sharply with increasing
galacto-centric distance. 

\begin{figure*}[t!]
\rotatebox{-90}{\epsfig{file=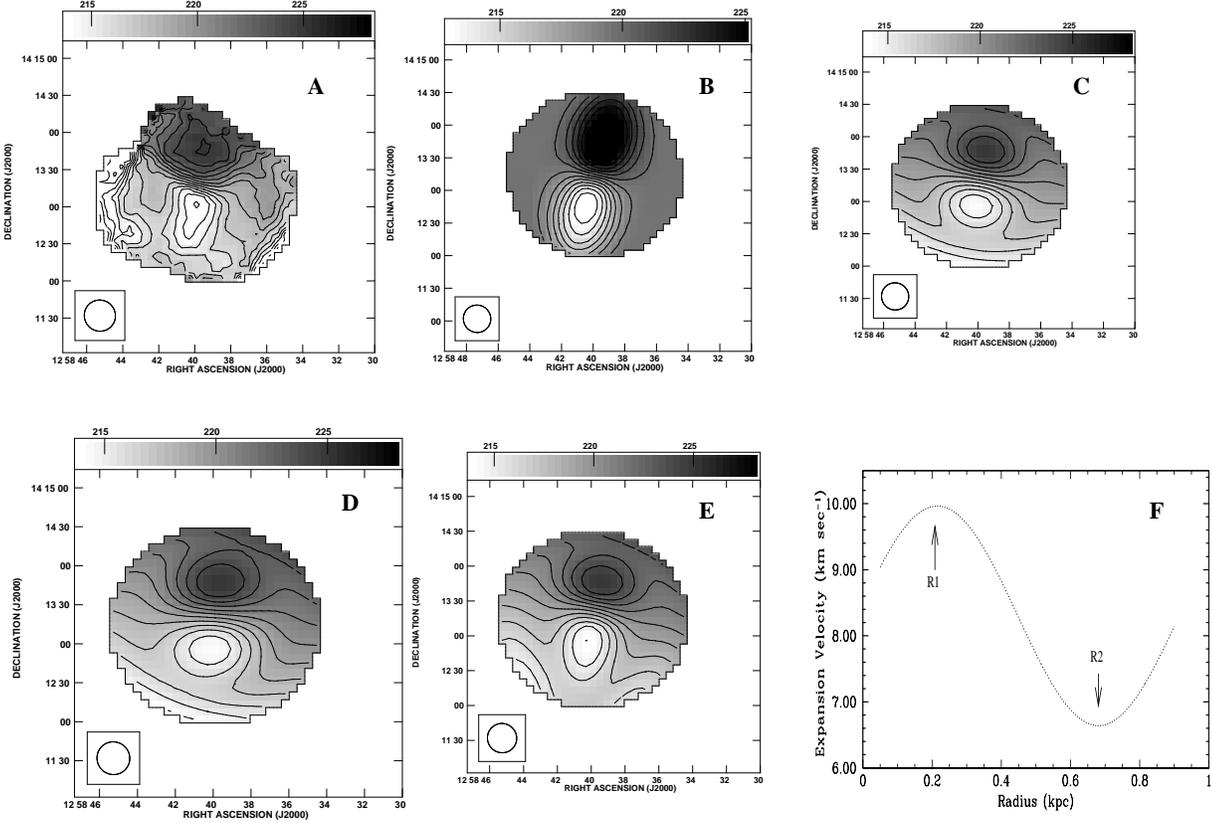,width=4.5truein}}
\caption{ [A] The $25{''} \times 25^{''}$ resolution moment~1 map.
              The contour levels go from $210$ to $225$~\kms in steps of 
	      1~\kms.
	  [B] The velocity field obtained using the rotation curve of
              \cite{carignan90}. Note that we have further assumed that
              the HI disk is intrinsically elliptical, with an axis
              ratio of 2:1. See the text for more details. The contours
              levels in this and succeeding panels are from $213$ to 
              $225$~\kms in steps of 1~\kms.
	  [C] The model velocity field with only expansion motion in the 
              gas. The expansion velocity  used to obtain the model is 
	      given in panel [F]. See also the discussion in the text.
          [D] The model velocity field with both  expansion and 
	      rotation motions in the gas. The expansion is the same
              as in panel [C]. The assumed rotation curve is linear
              and rises to a maximum of 6~\kms at the edge of the
              galaxy.  
          [E] The model velocity field  with rotational and asymmetric 
	      expansion motions. The rotational velocity is the same
              as used in panel [D], however the expansion curve (while
              similar in form to that in panels [C] and [D] has been
              scaled (in galacto-centric distance but not amplitude).
          [F] Expansion curve used to obtain the model velocity field. 
              See the text for more details.
            }
\label{fig:model}
\end{figure*}

	Our attempts to derive a rotation curve from our velocity field
were not successful. The errors in the estimated parameters were large,
as were the residuals between the model and the observed velocity
field. Our failure to find a good fit (as opposed to \cite{carignan90},
who were able to fit a rotation curve) is probably related to our
better sampling of the velocity field, which, as noted above, 
makes the misalignments and asymmetries in the velocity field more
striking. To provide a feel for the velocity field that would be produced
by circular rotation, we show in Fig.~\ref{fig:model}[B] the model  velocity
field that corresponds to the rotation curve of \cite{carignan90}. 
The disk has been taken to be intrinsically elliptical (with an
axis ratio of 2:1), so that despite having an inclination of $60^o$
(the inclination angle derived from the velocity field by 
Carignan et al. 1990) the projected model HI disk matches the fairly 
circular appearance of the observed HI disk. Essentially, the 
foreshortening along the kinematical minor axis is offset by 
the inherent ellipticity of the disk. As expected, although the model
produces closed isovelocity contours along the apparent morphological HI 
minor axis, the asymmetries seen in  the closed contours between  
northern and southern half of the galaxy, the kinks in the isovelocity 
contours towards the edges of the disk, as well as the offset between the 
kinematical and morphological center are not reproduced. As discussed
in Sect.~\ref{ssec:HI_Kin}, kinks in the outer isovelocity contours
can be produced by requiring the rotation curve to rise again, or 
by requiring  the outer parts of the disk to be extremely warped.
Quantitatively, to reproduce the observed kinks, the inclination angle 
is required to change by an amount sufficient to
cause the observed velocity at the edges to increase by a factor 
of $\sim$~2 compared to the unwarped model. Such extreme warps can,
in principle, lead to multiply peaked line profiles. However, because
of the low signal to noise ratio towards the edges, we cannot reliably
distinguish between single peaked and  multiply peaked line profiles in 
these regions. A more serious concern in modeling the velocity field of
GR8 as a rotating disk is the observed misalignment between the 
kinematical and HI major axes. As noted above, this requires the HI 
disk to be inherently elongated with an axis ratio of at least 2:1.
Such a highly non circular disk would be very unusual. Further, 
the inner regions of the galaxy (i.e. the distance at which the 
rotation curve of \cite{carignan90} peaks) will complete one rotation 
in $\sim$~80~Myr, while the rotation period at the edge of the disk
is $\sim$~1~Gyr. Hence, this differential rotation will wind up any
elongation in the disk on a timescale that is short compared to the
age of the galaxy.

	 Alternatively, as first proposed by \cite{lo93}, the observed 
velocity field of GR8  could also be the result of radial motions in 
the gas i.e expansion or contraction. Since the sign of the inclination
of the galaxy is unknown, it is not possible to distinguish between
inward and outward radial motions. Large scale bulk radial gas flows,
although difficult to understand in the context of normal spiral
galaxies, could nonetheless be plausible in small galaxies like GR8.
In models of dwarf galaxy formation and evolution, energy injected into
the ISM from stellar winds  and supernova explosions could drive 
significant expansive motions in the gas. In fact, in such models, 
dwarf galaxies below a critical halo  circular velocity of 
$\sim$ 100 \kms are expected to lose a significant fraction of their
ISM from the first burst of star formation (e.g. \cite{dekel86},
\cite{efstathiau00}). Expulsion of the ISM because of the energy
input from supernovae is also postulated as a possible mechanism 
for producing dwarf elliptical galaxies from gas rich progenitors
(e.g. Miralda-Escude \& Rees 1997). Observationally, outflows of ionized
material have been seen in star bursting dwarf galaxies 
(e.g. Marlowe et al. 1995). Of course, these models deal with the
expulsion of hot supernovae heated gas, where as, in this instance
we are dealing with cold neutral gas.  For sufficiently small galaxies
however, model calculations (Ferrara \& Tolstoy 2000) suggest that the ISM 
could be ``blown away'' i.e. that the ambient medium could be swept  out 
by the hot expanding supernovae superbubbles. This is in contrast to the
situation in slightly larger galaxies where there is instead a 
``blow out'' i.e. the supernovae heated hot gas pierces the ambient
disk material and escapes into the intergalactic medium. Although
a situation where the entire ISM is expanding outwards has not
yet been observed, expansion of the neutral ISM on smaller scales
has been observed in a number of starbursting dwarf galaxies.
Such expanding HI supershells have been seen in, for example, Holmberg~II 
(Puche et al. 1992), IC~2574 (Walter \& Brinks 1999) and Holmberg~I
(Ott et al. 2001). One should note however, that while the observational 
evidence for expanding shells in the ISM of these galaxies is 
reasonably good, the mechanism by which these shells have been 
created is less well established. \cite{stewart00} find that the giant
supershell in IC~2574 is probably driven by energy input 
from supernovae, while \cite{rhode99}, despite deep optical
imaging, do not find the star clusters that would be expected to be 
present in this scenario, at the centers of the HI holes in 
Holmberg~II 

	In light of the above discussion, and the ongoing star formation
in GR8, it may be reasonable to assume that there are large scale radial
flows in the galaxy. If we make this assumption, then the line of sight 
velocity $V_{\rm los}$ is related to the circular velocity $V_{\rm rot}$ 
and the radial velocity $V_{\rm exp}$ by the relation

\begin{equation}
V_{\mathrm{los}}=V_{\rm sys}+(V_{\rm rot}{\cos(\phi)}+
	        V_{\rm exp}\sin(\phi))\sin(i)
\end{equation}

where $V_{\rm sys}$ is the systemic velocity, $i$ is the inclination
angle, and $\phi$ is the azimuthal angle in the plane of the galaxy
($\phi = 0$ along the receding half of the kinematical major axis).
The simplest such model is one in which there is no rotation. 
Fig.~\ref{fig:model}[C] shows such a model for GR8. In this model 
the inclination angle of the disk is taken to be $20^o$ and 
the position angle $350^o$. These values were chosen to match the
observed velocity field, and are in good agreement with the
values expected from the ellipse fitting to the outer HI contours
(see Sect.~\ref{ssec:HI_dis}; note that in the case of radial
motion, the velocity gradient is maximum along the morphological 
minor axis and not the morphological major axis). The expansion is taken
to be centered on the kinematical center obtained from the velocity field,
and not the morphological center. Since radial motions are probably driven 
by energy from star formation, it is not necessary for the 
expansion center to be coincident with the geometric center of 
the HI disk. The expansion $V_{\rm exp}$ is assumed to be 
azimuthally symmetric, and its variation with galacto-centric 
distance is as shown in Fig.~\ref{fig:model}[F].
The rise in the expansion 
velocity  till the radius R1 produces the parallel isovelocity 
contours in the central regions of the galaxy, the fall
after R1 produces the closed contours. The rise in the expansion 
curve, from  radius R2  onwards, produces the kinks seen in the 
eastern and western edges of the velocity field. 
This particular form of expansion was chosen because it provides  
a good match to the observed velocity field. While it is possible 
that detailed gas dynamic modeling might be able to reproduce this 
curve, we have not attempted any such modeling in this paper.

	While a pure expansion model does produce the closed
contours along the morphological minor axis, it does not produce the
asymmetries in the velocity field noted in Sect.~\ref{ssec:HI_Kin}.
The next most natural model to try is hence one in which there is
also some rotation. A velocity field with the same $V_{\rm exp}$ as
before, but with non zero $V_{\rm rot}$ is shown in Fig.~\ref{fig:model}[D].
The rotation curve has been assumed to be linear; it
rises to a maximum of 6~\kms at the edge of the galaxy.
A linearly rising rotation curve was chosen because this
form of  rotation curve is typical of dwarf galaxies. 
Other types of rotation curves, i.e.  a  constant rotation curve, 
a Brandt and an exponential curve (which are seldom observed for 
dwarf galaxies) were also tried. While a constant rotation curve
gives a poor fit to the data, Brandt and exponential curves do
not provide a better fit to the observed velocity field than that
provided by a linear curve. Since these curves  introduce many 
more free parameters in the model without improving the fit quality
they  were not explored further.
The rotation is assumed to be centered on the morphological center of the galaxy. 
The inclination and position angle are the same as for the previous 
model. As can be seen, this does reproduce the asymmetry in the kinks
in the isovelocity contours between the eastern and western
halves of the galaxy. However, it still does not reproduce the asymmetries in
the closed contours. A model which does reproduce most of the features
of the observed velocity field is shown in Fig.~\ref{fig:model}[E]. 
This model is similar to that used to produce Fig.~\ref{fig:model}[D],
the difference being that the expansion curve is no longer assumed
to be azimuthally symmetric. The positions of  R1 and R2 in the 
expansion curve (see Fig.~\ref{fig:model}[F]) were allowed to be different
at different azimuthal angles
in the southern half of the galaxy. However, the maximum amplitude of 
the expansion curve was taken to be the  same in all 
azimuthal directions. The effect of this was to reproduce the 
elongated closed contours in the southern half of the galaxy. 
This asymmetry between the kinematics in the northern and southern 
halves may be related to the corresponding asymmetry in the distribution 
of HII regions (see Fig.~\ref{fig:hii}, and also the discussion
below). The match can obviously 
be improved by also allowing an azimuth angle dependent 
scaling of the amplitude of the expansion curve, but in the absence of a 
physically motivated prescription for the scaling factor, this would 
not add much to our understanding of the galaxy's kinematics. 
It should be noted that it has not been shown that our chosen 
model provides a unique (or even ``best'' in some rigorous statistical
sense) fit to the observed kinematics of the galaxy. It is possible
that one could find different forms for the expansion and rotation
curves which also provide adequate fits to the observed velocity field.
Strictly speaking, a more robust method would have been to determine a
least squares fit to the observed velocity field, allowing for both
expansion and rotation. This approach has however not been attempted
in this paper.

	Fig.~\ref{fig:model}[E] shows that the observed velocity
field of GR8 can be quite well matched by a combination of rotational
and expansion motions.  Assuming that this interpretation is correct,
the natural question that arises is, what drives the expansion
of the gas?  Energy input from star formation is the obvious
suspect. For an expansion velocity of $\sim 10$~\kms and an HI 
mass of $\sim~10^7M_\odot$ the corresponding kinetic energy is
$\sim~ 10^{52}$ erg. If we assume that the kinetic energy imparted 
to the ISM by one supernova explosion is $\sim~ 10^{51}$ erg 
(e.g. Reynolds 1988),  this implies that kinetic energy required for
the expansion motion is equivalent to the energy output of $\sim 10$ 
supernovae. It is plausible that this number of supernovae have occurred in
GR8 in the recent past. The lack of detection of radio continuum sources
(corresponding to the supernovae remnants) would then place a limit
on the magnetic field in the galaxy. On the other hand, no star 
clusters are located at the center of expansion.
However, as can be seen from Fig.~\ref{fig:hii}, the majority 
of the HII regions associated with the three HI clumps lie 
on the inner edges of the clumps, (i.e. towards the center of 
the galaxy).  In a study of HII regions in dwarf galaxies \cite{elmegreen00}
found that the HII regions tend to have
a higher pressure than the average pressure in the disk. They
suggest that the HII regions could still be in pressure equilibrium
if they preferentially lie in dense HI clumps, where the ambient
pressure is higher than the average over the disk. Of the sample
of galaxies studied by \cite{elmegreen00} GR8 showed the largest
pressure anomaly; the pressure in the HII regions was found to
be at least a factor of $\sim$~55 times larger than the average
disk pressure. Since these HII regions tend to lie at the inner
edges of the HI clumps, this over pressure could possibly drive
the clumps outwards. It is interesting to note in this regard,
that the star formation history of these clumps indicates that
they have been forming stars continuously over at least  
last 500 Myr, i.e. the clumps themselves are gravitationally
bound (Dohm-Palmer et al. 1998). The measured expansion velocity 
($\sim 10$\kms) is also considerably smaller than the escape velocity 
(which would be $\sim 30$\kms, if we assume that GR8 is dark matter 
dominated and has a dynamic mass to light ratio $\sim 10$), which 
means that the neutral ISM is still bound. This is consistent with the
models of dwarf galaxy evolution which include a clumpy ISM-- in such
models the cold clumpy material does not escape from the galaxy 
\cite{andersen00}.

	So far we have been treating the radial motions as
expansion. Since the sign of the radial motion is unconstrained,
we should also note that the velocity field could instead arise
from infall. In this case, gas is swirling inwards into the
galaxy. The model would then be that GR8 is forming from
the merger of the three clumps, and that the diffuse gas
and stars are material that has be tidally stripped from the
clumps and which is now settling down to form a disk. However,
in this scenario, it is unclear if one would obtain the observed
velocity field, which doesn't show any clear signature of tidal
interaction. Another possible infall scenario is that the gas is now
falling back towards the center of the potential after a previous 
expansion phase.

	To conclude, we have presented  deep, high velocity resolution 
($\sim 1.6$ km sec$^{-1}$) GMRT HI 21cm synthesis images for the faint 
($M_B \sim -12.1$) dwarf irregular galaxy GR8. We find that though the HI 
distribution in the galaxy is very clumpy, there is nonetheless
substantial diffuse gas. The velocity field of the galaxy is not chaotic,
but shows a systematic large scale pattern. We are unable to fit this
pattern with either pure rotation or pure expansion. From an inspection
of the velocity field however, the following qualitative remarks can
be made. If this pattern is treated as arising because of rotation, 
then (i)~the rotation curve would have to be sharply falling, and the
disk would have to be extremely warped at the outer edges and (ii)~the disk
has to inherently elliptical, with an axis ratio $\simgeq \ 2$. Such a 
disk would get quickly wound up due to differential rotation. For these
reasons we regard it unlikely that GR8's velocity field is due to
pure rotation. A more likely model is one in which the kinematics
of GR8 can be described as a combination of radial and circular motions. 
Such a model provides a reasonable fit to the observed velocity field.
In this interpretation, in case the radial motions are outwards, then 
they could be driven by  the star formation in GR8; a previous study 
(Elmegreen \& Hunter 2000) has shown that the pressure in the HII regions in 
this galaxy is at least $55$ times greater than the average pressure 
in the disk. The measured expansion velocity is considerably less than
the estimated escape velocity, so even in this interpretation the cold
gas is still bound to the galaxy. Finally, the radial motions could 
also be interpreted as infall, in which case GR8 is either in the 
process of formation, or the ISM is falling back after a previous 
phase of expansion.

\begin{acknowledgements}
        The observations presented in this paper would not have been 
possible without the many years of dedicated effort put in by the
GMRT staff in order to build the telescope. The GMRT is operated 
by the National Centre for Radio Astrophysics of the Tata Institute 
of Fundamental Research. We are grateful to Rajaram Nityananda for 
many helpful conversations and valuable comments. 
\end{acknowledgements}

\end{document}